\documentclass[aps,prl,twocolumn,superscriptaddress,showpacs,preprintnumbers,amsmath,amssymb]{revtex4}
\usepackage{graphicx} 
\usepackage{dcolumn}  
\usepackage{xspace}
\usepackage{color}
\graphicspath{{ps}}

\newcommand{\bb}{\ensuremath{B \overline{B}}\xspace}

\newcommand{\btoetapi}{\ensuremath{B^\pm \to \eta \pi^\pm}\xspace}
\newcommand{\btoetak}{\ensuremath{B^\pm \to \eta K^\pm}\xspace}

\newcommand{\btoetakz}{\ensuremath{B^0 \to \eta K^0}\xspace}

\newcommand{\etagg}{\ensuremath{\eta_{\gamma \gamma}}\xspace}
\newcommand{\etapi}{\ensuremath{\eta_{3\pi}}\xspace}

\newcommand{\de}{\ensuremath{\Delta E}\xspace}
\newcommand{\br}{\ensuremath{\mathcal{B}}\xspace}
\newcommand{\acp}{\ensuremath{A_{CP}}\xspace}

\def\calL{{\mathcal L}}

\def\Mbc{M_{\rm bc}}
\def\LR{{\mathcal R}}

\begin{document}


\preprint{\vbox{ 
}}

\title{ \quad\\[1.0cm] \boldmath Improved Measurements of Branching Fractions 
and $CP$ Asymmetries in $B\to \eta h$ Decays}


\affiliation{Budker Institute of Nuclear Physics, Novosibirsk}
\affiliation{Chiba University, Chiba}
\affiliation{Chonnam National University, Kwangju}
\affiliation{University of Cincinnati, Cincinnati, Ohio 45221}
\affiliation{The Graduate University for Advanced Studies, Hayama, Japan}
\affiliation{University of Hawaii, Honolulu, Hawaii 96822}
\affiliation{High Energy Accelerator Research Organization (KEK), Tsukuba}
\affiliation{Hiroshima Institute of Technology, Hiroshima}
\affiliation{Institute of High Energy Physics, Chinese Academy of Sciences,
Beijing}
\affiliation{Institute of High Energy Physics, Vienna}
\affiliation{Institute of High Energy Physics, Protvino}
\affiliation{Institute for Theoretical and Experimental Physics, Moscow}
\affiliation{J. Stefan Institute, Ljubljana}
\affiliation{Kanagawa University, Yokohama}
\affiliation{Korea University, Seoul}
\affiliation{Kyungpook National University, Taegu}
\affiliation{Swiss Federal Institute of Technology of Lausanne, EPFL, Lausanne}
\affiliation{University of Maribor, Maribor}
\affiliation{University of Melbourne, Victoria}
\affiliation{Nagoya University, Nagoya}
\affiliation{Nara Women's University, Nara}
\affiliation{National Central University, Chung-li}
\affiliation{National United University, Miao Li}
\affiliation{Department of Physics, National Taiwan University, Taipei}
\affiliation{H. Niewodniczanski Institute of Nuclear Physics, Krakow}
\affiliation{Nippon Dental University, Niigata}
\affiliation{Niigata University, Niigata}
\affiliation{University of Nova Gorica, Nova Gorica}
\affiliation{Osaka City University, Osaka}
\affiliation{Osaka University, Osaka}
\affiliation{Panjab University, Chandigarh}
\affiliation{Peking University, Beijing}
\affiliation{RIKEN BNL Research Center, Upton, New York 11973}
\affiliation{University of Science and Technology of China, Hefei}
\affiliation{Seoul National University, Seoul}
\affiliation{Shinshu University, Nagano}
\affiliation{Sungkyunkwan University, Suwon}
\affiliation{University of Sydney, Sydney NSW}
\affiliation{Toho University, Funabashi}
\affiliation{Tohoku Gakuin University, Tagajo}
\affiliation{Tohoku University, Sendai}
\affiliation{Department of Physics, University of Tokyo, Tokyo}
\affiliation{Tokyo Institute of Technology, Tokyo}
\affiliation{Tokyo Metropolitan University, Tokyo}
\affiliation{Tokyo University of Agriculture and Technology, Tokyo}
\affiliation{Virginia Polytechnic Institute and State University, Blacksburg,
Virginia 24061}
\affiliation{Yonsei University, Seoul}
 \author{P.~Chang}\affiliation{Department of Physics, National Taiwan University, Taipei} 
  \author{K.~Abe}\affiliation{Tohoku Gakuin University, Tagajo} 
  \author{I.~Adachi}\affiliation{High Energy Accelerator Research Organization
(KEK), Tsukuba} 
  \author{H.~Aihara}\affiliation{Department of Physics, University of Tokyo,
Tokyo} 
  \author{D.~Anipko}\affiliation{Budker Institute of Nuclear Physics,
Novosibirsk} 
  \author{A.~M.~Bakich}\affiliation{University of Sydney, Sydney NSW} 
  \author{E.~Barberio}\affiliation{University of Melbourne, Victoria} 
  \author{A.~Bay}\affiliation{Swiss Federal Institute of Technology of
Lausanne, EPFL, Lausanne} 
  \author{U.~Bitenc}\affiliation{J. Stefan Institute, Ljubljana} 
  \author{I.~Bizjak}\affiliation{J. Stefan Institute, Ljubljana} 
% ...
\collaboration{The Belle Collaboration}

\begin{abstract}
We report improved measurements of $B$ decays 
with an $\eta$ meson in the final state using $492 \mathrm{~fb}^{-1}$ of 
data collected by the Belle detector at the KEKB $e^+ e^-$ collider.  
We observe the decays  $\btoetapi$ and   $\btoetak$ and measure the 
branching fractions 
$\br(\btoetapi) = ( 4.2\pm 0.4 \mathrm{(stat)} \pm 0.2
 \mathrm{(sys)}) \times 10^{-6}$ and
$\br(\btoetak) = ( 1.9\pm 0.3 \mathrm{(stat)}^{+0.2}_{-0.1}
\mathrm{(sys)}) \times 10^{-6}$. The corresponding $CP$-violating
asymmetries are measured to be $-0.23\pm 0.09 \mathrm{(stat)} 
\pm 0.02 \mathrm{(sys)}$ for $\eta \pi^\pm$ and $-0.39\pm 0.16 \mathrm{(stat)} 
\pm 0.03 \mathrm{(sys)}$ for $\eta K^\pm$. 
We also search for $B^0\to \eta K^0$ decays and set an 
 upper limit of $1.9 \times 10^{-6}$ at the 90\% confidence level. 

\end{abstract}


\maketitle

\tighten

{\renewcommand{\thefootnote}{\fnsymbol{footnote}}}
\setcounter{footnote}{0}
Charmless $B$ decays provide a rich sample to understand  $B$ decay
dynamics and to search for $CP$ violation. 
The decay $B\to \eta K$ proceeds through a $b\to s$ penguin process and a 
$b\to u$ tree transition. Interference from the two penguin 
processes, $b\to s \overline s s$ and  $b\to u \overline u s$, and the known
$\eta-\eta^\prime$ mixing are expected to enhance the 
$B\to \eta^\prime K$ branching 
fraction but suppress $B\to\eta K$ \cite{lipkin}. The situation is reversed 
 for the $\eta K^*$ and $\eta^\prime K^*$ modes since the $\eta$ and  
$K^*$ mesons are  
in a relative p-wave rather than an s-wave state.
Experimental results \cite{etaprime, etahbelle, etahbabar,etakstbabar} have 
confirmed this picture but more precise measurements of $\eta K$ and 
$\eta^\prime K^*$ are needed for a  quantitative understanding of the 
underlying dynamics. Moreover, the  penguin amplitude 
of $\eta K$ may interfere with the $b\to u$ amplitude, resulting in a large 
direct $CP$ asymmetry ($A_{CP}$) \cite{soni}. Theoretical expectations for
contributions from other mechanisms \cite{hard,singlet,scet} also suggest a 
large $A_{CP}$ although the sign could be positive or negative.
Our earlier measurements with limited statistics
\cite{etahbelle} indicated a large negative $A_{CP}$ central value for 
$\eta K^\pm$.

The study of $B^0\to \eta K^0$ is of particular interest because this decay
is a $CP$ eigenstate and could be used for time dependent $CP$ measurements. 
The dominant process in  $B\to \eta \pi$ decays is the (external) 
$b\to u$ tree while a suppressed $b\to d$ penguin process may also contribute. 
It has been argued \cite{theory1,hard} that the direct $CP$ violating 
asymmetry could be large in
the $\eta \pi^\pm$ and $\eta^\prime \pi^\pm$ modes, whose branching fractions 
are expected to be around $(2-5)\times 10^{-6}$ \cite{scet,theory1}.

In this paper, we report improved  measurements of branching fractions and 
partial rate asymmetries for  $B\to\eta h$ decays, where $h$ is a 
charged or neutral $K$ meson or a charged $\pi$  meson. 
The partial rate asymmetry for
charged $B$ decays is  defined to be:
\begin{eqnarray}
\acp=\frac{N( B^- \to \eta h^-)-N(B^+ \to \eta h^+)}
{N(B^- \to \eta h^-)+N(B^+ \to \eta h^+)},
\end{eqnarray}
where $N(B^-\to \eta h^-)$ is the yield obtained for the 
$B^- \to \eta h^-$ decay and
$N(B^+\to \eta h^+)$ denotes that of the charge-conjugate mode.
 The data sample consists of 535
million \bb pairs (492 fb$^{-1}$) collected
with the Belle detector at the KEKB $e^+e^-$ asymmetric-energy  
(3.5 on 8~GeV) collider~\cite{KEKB} operating at the $\Upsilon(4S)$ resonance.
Throughout this paper, the inclusion of the charge-conjugate  decay mode is 
implied unless otherwise stated.

The Belle detector is a large-solid-angle magnetic
spectrometer that
consists of a silicon vertex detector (SVD),
a 50-layer central drift chamber (CDC), an array of
aerogel threshold \v{C}erenkov counters (ACC),
a barrel-like arrangement of time-of-flight
scintillation counters (TOF), and an electromagnetic calorimeter (ECL)
comprised of CsI(Tl) crystals located inside
a superconducting solenoid coil that provides a 1.5~T
magnetic field.  An iron flux-return located outside 
the coil is instrumented to detect $K_L^0$ mesons and to identify
muons (KLM).  The detector
is described in detail elsewhere~\cite{Belle}.
In August 2003, the three-layer SVD was replaced by a four-layer device 
with greater radiation tolerance ~\cite{SVD2}. The data sample used in this 
analysis consists of 
140 fb$^{-1}$ of data with the old SVD (Set I) and 
352 fb$^{-1}$ with the new one (Set II).

The event selection and $B$ candidate reconstruction are similar to 
those documented in our previous publication \cite{etahbelle}. 
Two $\eta$ decay channels are considered in this analysis: 
$\eta\to \gamma\gamma$ ($\eta_{\gamma\gamma}$) and $\eta\to \pi^+\pi^-\pi^0$
 ($\etapi$). We require photons from the $\eta$ and $\pi^0$ candidates to have 
laboratory energies ($E_{\gamma}$)
above 50 MeV. In the $\etagg$ reconstruction, the 
photon energy asymmetry, $\frac{|E_{\gamma 1}-E_{\gamma 2}|}{
   E_{\gamma 1}+E_{\gamma 2}}$, is required to be  
less than 0.9 to reduce the large combinatorial background from soft photons.
 Neither photon from $\eta_{\gamma\gamma}$ is allowed
to pair with any other photon having $E_\gamma > 100$ MeV  to form a 
$\pi^0$
candidate. Candidate $\pi^0$ mesons are selected by requiring the
two-photon invariant mass to be in a mass window between 115 MeV/$c^2$ and
152 MeV/$c^2$.  The momentum vector of each photon is then readjusted to
constrain the mass of the photon pair to the nominal $\pi^0$ mass.

Candidate $\etapi$ mesons are reconstructed by combining  $\pi^0$ candidates 
with at
least 250 MeV/$c$ laboratory momentum with a pair of oppositely charged 
tracks that 
originate from the interaction point (IP). 
We impose the following requirements on the invariant mass of 
the $\eta$ candidates in both data sets: 
516 MeV/$c^2 < M_{\gamma\gamma} < 569$ MeV/$c^2$ for 
$\etagg$ and 539 MeV/$c^2 < M_{3\pi} <556$ MeV/$c^2$ for $\etapi$. After the 
selection of each candidate, the $\eta$ mass constraint is implemented  
by readjusting the momentum vectors 
of the daughter particles.   

Charged tracks are required to come from the IP.  Charged kaons and pions,
which are combined with $\eta$ mesons to form $B$ candidates,
are identified using a $K(\pi)$  likelihood $L_K(L_\pi)$ obtained 
by combining information from the CDC ($dE/dx$),
the TOF and the ACC. Discrimination between kaons and pions is achieved 
through a  requirement on the likelihood ratio $L_{K}$/($L_{\pi}+L_{K}$). 
Charged tracks with  
likelihood ratios greater than 0.6 are regarded as kaons, and less than 0.4
as pions. Charged tracks that are positively identified as 
electrons or muons are rejected. The $K/\pi$ identification efficiencies (PID)
and misidentification rates are determined from a sample of 
$D^{*+}\to D^0\pi^+, D^0\to K^-\pi^+$ decays with kaons and pions in the
same kinematic region of two-body $B$ decays.  
 The kaon (pion) identification efficiency is 83\% (90\%) and 6.4\% (11.7\%)
of pions (kaons) will be misidentified as kaons (pions).
The systematic error of the $K/\pi$ selection is about 1.3\% for pions and 1.5\% for kaons, respectively.

$K^0_S$ candidates 
are reconstructed from  pairs of oppositely-charged tracks
with an invariant mass ($M_{\pi\pi}$) between 480 MeV/$c^2$ and 516 MeV/$c^2$.
Each candidate must have a displaced vertex with a flight direction
consistent with that of a $K^0_S$-meson originating from the IP.

Candidate $B$ mesons are identified using the beam-energy constrained mass,
$\Mbc =  \sqrt{E^2_{\mbox{\scriptsize beam}} - P_B^2}$,
and the energy difference, $\Delta E = E_B  - E_{\mbox{\scriptsize beam}}$,
where $E_{\mbox{\scriptsize beam}}$ is the run-dependent beam energy in the
$\Upsilon(4S)$ rest frame  determined  from
$B\to D^{(*)}\pi$ events,
and $P_{B}$ and $E_B$ are the momentum and energy, respectively of  the
$B$ candidate in the $\Upsilon(4S)$ rest frame.
 The resolutions in $\Mbc$ and 
$\de$ are about 3 MeV/$c^2$ and 20--30 MeV, respectively.  
Events with $\Mbc >5.2$ GeV/$c^2$ and $|\de|<0.3$ GeV are selected for the 
 analysis.   

The dominant background comes from the $e^+e^-\rightarrow q\bar{q}$ continuum, 
where $q= u, d, s$ or $c$. To distinguish signal from the jet-like continuum 
background,  event shape variables and $B$ flavor tagging information 
are employed. We combine the correlated shape variables into a 
Fisher discriminant \cite{fisher} and then compute a likelihood that is the
 product of probabilities based on this discriminant and $\cos\theta_B$, where $\theta_B$ is 
the angle between the $B$ flight direction and the beam direction
in the $\Upsilon(4S)$ rest frame. A likelihood ratio, 
$\LR = {\calL}_s/({\calL}_s + {\calL}_{q \bar{q}})$, is formed from signal
(${\calL}_s$) and background (${\calL}_{q \bar{q}}$) likelihoods, obtained 
from  Monte Carlo simulation (MC) and from data with 
$\Mbc< 5.26$ GeV/$c^2$, respectively. Signal MC events for the charged $B$ 
modes are generated with the PHOTOS \cite{photos} simulation package  to take 
into account  final state radiation.  Additional 
background discrimination is provided by $B$ flavor tagging. Events that 
contain a lepton (such as those used in high quality tagging) are more 
likely to be 
$B \overline B$ events so a looser $\LR$ requirement is applied. 
The standard Belle $B$ tagging package \cite{tagging} provides two outputs: 
a discrete variable ($q$) indicating the tagged side flavor and a dilution 
factor ($r$)
ranging from zero for no flavor information to unity for unambiguous
flavor assignment. Since the charged $B$ modes are flavor specific, the wrong
flavor tagged events are likely to be background and a tight $\LR$ requirement
can be applied. We divide the data into six sub-samples based on 
the $q$ and $r$ information  for the charged modes and the $r$ value only 
for the neutral mode.
Continuum suppression is achieved by applying a mode-dependent requirement on 
$\LR$ for events in each sub-sample  that maximizes  
$N_s^{\rm exp}/\sqrt{N_s^{\rm exp}+N_{q\bar{q}}^{\rm exp}}$,
where $N_s^{\rm exp}$ is the number of signal events expected  from MC and
$N_{q\bar{q}}^{\rm exp}$ denotes the number of background events estimated from data. After applying the $\LR$ requirements, we select
one candidate per event based on the best $\LR$. The fraction of events
with multiple candidates are $\sim 1\%$ for the $\gamma\gamma$ mode and
$\sim$2-3\% for the $\pi^+\pi^-\pi^0$ mode.

Using a large MC sample, all other backgrounds are found to be negligible 
 except for $\eta K^+ (\eta \pi^+)$ reflecting into the $\eta \pi^+ (\eta K^+)$
sample, due to $K^+\to \pi^+ (\pi^+\to K^+)$ 
misidentification, and the feed-down from charmless $B$ decays, predominantly  
$B\to \eta K^*(892)$ and  $B\to \eta\rho(770)$. 
We include the reflection and charmless components in the fit used 
to extract the signal.   

The signal yields and partial rate asymmetries
are obtained using an extended 
unbinned maximum-likelihood (ML) fit with input variables $\Mbc$ and $\de$.
The likelihood is defined as: 
\begin{eqnarray} 
\mathcal{L} & = & e^{-\sum_j N_j}
\times \prod_i (\sum_j N_j \mathcal{P}^i_j) \;\;\; \mbox{and} \\
\mathcal{P}^i_j & = &\frac{1}{2}[1- q^i \cdot \acp{}_j ]
P_j(M^i_{\rm bc}, \Delta E^i),
\end{eqnarray}
where $i$ is the identifier of
the $i$-th event and $N_j$ is the number of events for the category $j$, 
which corresponds to either signal, $q\bar{q}$ continuum,
the reflection due to $K$-$\pi$ misidentification, or
background from other charmless $B$ decays. 
$P_j(M_{\rm bc}, \Delta E)$ is the two-dimensional probability
density function (PDF) in
$M_{\rm bc}$ and $\Delta E$, and $q$ indicates the $B$ meson flavor,
$B^+(q=+1)$ or $B^- (q=-1)$.
For the neutral $B$ mode, 
$\mathcal{P}^i_j$ in Eq. 2 is simply 
$P_j(M_{\rm bc}^i, \Delta E^i)$ and there is no  component from charged 
particle misidentification.

In Ref. \cite{acpkpi} we reported that in both data sets the PID efficiency
is slightly different for positively and negatively charged particles. 
Therefore, the raw asymmetry defined in Eq. 1 
must be corrected. This efficiency 
difference results in an $A_{CP}$ bias of  $-0.005$ (+0.005) for $\eta \pi$ 
($\eta K$). The bias is subtracted from 
 the raw asymmetry.

\begin{table*}[th]
\caption{ Detection efficiency ($\epsilon$) including sub-decay branching 
fraction \cite{pdg}, yield,
significance (Sig.), measured branching fraction ($\mathcal{B}$), 
the 90\% C.L. upper limit (UL) and $A_{CP}$ for the $B\to \eta h$ decays. 
The first errors
in columns 3, 5 and 7 are statistical and the second errors are systematic.}

\begin{tabular}{lcccccc} \hline\hline
Mode & $\epsilon(\%)$ &    Yield & \hspace{0.2cm}Sig. 
\hspace{0.2cm} & $\mathcal{B} (10^{-6})$  & UL$(10^{-6}$)& $A_{CP}$ \\
\hline
\hline
 \btoetapi & 
        & & 14.7& $4.2\pm 0.4\pm 0.2$ &
         & $-0.23\pm 0.09 \pm 0.02$
        \\
 \hspace{0.2cm} $\etagg \pi^\pm$  & 8.3 &$183\pm 20$ 
& 11.9 &$4.1^{+0.5}_{-0.4}\pm 0.2$ & &$-0.11\pm 0.11 \pm 0.01$ \\
\hspace{0.2cm}$\eta_{3\pi} \pi^\pm$ & 3.1 & $73^{+13}_{-12}$ &
8.7& $4.4^{+0.8}_{-0.7} \pm 0.3$ & & 
$-0.52\pm 0.16 \pm 0.02$ \\
 \btoetak &
      &  & 8.1 & $1.9\pm 0.3^{+0.2}_{-0.1}$ &  & $-0.39\pm 0.16\pm 0.03$ \\
\hspace{0.2cm}$\etagg K^\pm$ & 7.3 & $72^{+14}_{-13}$ 
              & 6.4 & $1.9^{+0.4}_{-0.3}\pm 0.1$& & $-0.30\pm 0.19\pm 0.02$  \\
\hspace{0.2cm}$\eta_{3\pi} K^\pm$  & 2.7 & $29\pm 8$ 
         & 4.8 & $2.0^{+0.6}_{-0.4}\pm 0.2$ & & $-0.55^{+0.27+0.05}_{-0.28-0.04} $ \\         
 $B^0\to \eta K^0$ &
       & & 2.9 & $1.1\pm 0.4\pm 0.1$ & $<1.9$ & \\
\hspace{0.2cm}$\etagg K^0$ &2.6 &$16\pm 8$ & 2.6 & $1.1^{+0.6}_{-0.5}\pm 0.1$ &$<2.2$ & \\
\hspace{0.2cm}$\eta_{3\pi} K^0$ &1.0 &$4.6^{+4.6}_{-3.7}$ & 1.2 & $0.9^{+0.9}_{-0.7}\pm 0.1$ &$<2.4$  & \\
\hline
\end{tabular}
\label{tab:result}
\end{table*}

The PDFs for the signal, the reflection background and the 
charmless feed-down are modeled with two-dimensional 
$\Mbc$-$\de$ smooth functions obtained using large MC samples.
 The signal peak positions and resolutions in $\Mbc$ and $\de$
are adjusted according to the data-MC differences using large control samples 
of $B\to D\pi$  and $\overline{D}{}^0\to K^+\pi^-\pi^0$ decays. 
The continuum background in $\de$ is described by a first- or second-order
polynomial, while the $\Mbc$ distribution is parameterized by an 
ARGUS function, $f(x) = x \sqrt{1-x^2}\;{\rm exp}\;[ -\xi (1-x^2)]$, where 
$x$ is $\Mbc/E_{\rm beam}$ \cite{argus}. 
The continuum PDF is thus formed by the product of an ARGUS function and a 
polynomial, where $\xi$ and the coefficients of the polynomial are free 
parameters. 

The partial rate asymmetries of the charmless $B$ backgrounds are fixed to zero 
in the 
fit while the $A_{CP}$ and  normalizations of the reflection components 
are fixed to expectations based on the $B^+\to \eta K^+$ and 
$B^+\to \eta \pi^+$  partial rate 
asymmetries and branching fractions, as well as   
$K^+\leftrightarrow \pi^+$ fake rates. The reflection yield and $A_{CP}$ 
are first input with the assumed values and are then
recalculated according to our measured results.

Table~\ref{tab:result} shows the measured branching
 fractions for each decay mode
as well as  other quantities associated with the measurements.
The efficiency for each mode is determined using  MC simulation and
corrected for the data-MC discrepancy obtained from the control sample studies. 
In addition to the  particle identification performance discrepancy,  
 our MC slightly overestimates the 
efficiency for detecting low momentum $\pi^0$s, which results in a 3.1\% correction
for the $\etapi$ mode. The combined branching fraction for the two datasets
is computed as the sum of the yield divided by its efficiency in each set
divided by the number of $B$ mesons, while the partial rate asymmetry
for the charged mode is computed using the sum of the yield divided by its
efficiency in each set in Eq. 1.  
The combined branching fraction and partial rate asymmetry of the two $\eta$ 
decay modes are obtained
from the weighted average assuming the errors are Gaussian. 
The number of $B^+B^-$ and $B^0\overline{B}{}^0$ pairs are assumed to be
equal.
 Figure \ref{fig:mbde} shows the $\Mbc$ and $\de$ projections after requiring
events to satisfy $-0.10$ GeV $<\de<0.08$ GeV and $\Mbc>5.27$ GeV/$c^2$, 
respectively. 

\begin{figure}[htb]
\includegraphics[width=0.48\textwidth]{mbde.epsi}
\caption{$\Mbc$ and $\de$ projections for (a,b) $\btoetapi$, (c,d) $\btoetak$,
and (e,f) $\btoetakz$ decays with 
the $\etagg$ and $\etapi$ modes combined. Open
histograms are data, solid curves are the fit functions, dashed lines show
the continuum contributions and shaded histograms are the feed-down component 
from charmless $B$ decays.  
The small contributions around $\Mbc = 5.28$ GeV/$c^2$ and 
$\de =\pm 0.05$
GeV in (a)-(d) are the reflection backgrounds from $\btoetak$ and $\btoetapi$.  
}
\label{fig:mbde}
\end{figure}

Systematic uncertainties  due to the signal PDFs used in the fit
are estimated by performing the fit after varying the signal peak positions and 
resolutions by one standard deviation ($\sigma$). We also examine the changes in yield and $A_{CP}$ when the requirement of no asymmetry for the charmless 
background is removed.
In $B^\pm \to \eta \pi^\pm$,
the reflection yields  are estimated to be $9.4\pm 3.1$ events 
for the $\etagg$ mode and $3.6 \pm 1.9$ for $\etapi$ while in 
$B^\pm \to \eta K^\pm$, the reflection yields are $13.9\pm 3.7$ for $\etagg$ and
$4.6\pm 2.1$ for $\etapi$.  The reflection yields and their $A_{CP}$ values 
 are varied by one standard deviation in the fit
to obtain the corresponding systematic uncertainties.
The quadratic 
sum of the deviations from the central value gives the systematic 
uncertainty in the fit. A statistical significance 
is calculated as ${\cal S} = \sqrt{-2\ln\calL_0 - 
 (-2\ln\calL_{\rm max})}$, where $-2\ln\calL_0$ is for zero signal yield and  
$-2\ln\calL_{\rm max}$ is for the best-fit value. The final significance 
including systematic uncertainty is taken as ${\cal S} = {\cal S}_o - 
\sqrt{\Sigma ({\cal S}_o -{\cal S}_i})^2$, where ${\cal S}_o$ is the 
statistical significance for the fit and ${\cal S}_i$ is the significance 
obtained for each systematic check with the value smaller than ${\cal S}_o$.

\begin{table*}[th]
\caption{The systematic uncertainties for the $B\to \eta h$ branching fractions, given in \%. The fit systematic errors include the uncertainties due to the 
 signal PDFs, the yields of the reflection backgrounds and the 
 partial rate asymmetries of the charmless $B$ and  reflection backgrounds. 
 }

\begin{tabular}{lcccccc} \hline\hline
Sources  & $ \etagg \pi^\pm$ &$ \etapi \pi^\pm$ & 
 $  \etagg K^\pm$ & $ \etapi K^\pm$ & 
 $ \etagg K^0$ & $\etapi K^0$ \\ \hline
Fit & $^{+2.6}_{-3.0}$ & $^{+3.0}_{-3.6}$ & $^{+6.1}_{-4.3}$ &
      $^{+6.2}_{-6.4}$ & $^{+6.0}_{-6.2}$ & $\pm 6.7$ \\
$\cal R$ requirement & $\pm 1.2$& $\pm 1.2$ & $\pm 1.2$ & $\pm 1.6$ & 
       $\pm 1.4$ & $\pm 1.4$ \\
Tracking & $\pm 1.0$ & $ \pm 3.0$ & $\pm 1.0$ & $\pm 3.0$ & $-$ & $\pm 2.0$ \\
PID & $\pm 1.3$ & $\pm 1.3$ & $ \pm 1.5$ & $\pm 1.5$ & $-$ & $-$ \\  
$K^0_S$ reconstruction & $-$ & $-$ & $-$ &$-$ & $\pm 4.9$ & $\pm 4.9$ \\
$\gamma\gamma$ reconstruction & $\pm 4.0$ & $\pm 4.0$ & $\pm 4.0$ & $\pm 4.0$ 
 &$\pm 4.0$ & $\pm 4.0$ \\
${\cal B}(\eta\to \gamma\gamma)$ & $\pm 0.7$ & $-$ & $\pm 0.7$ & $-$ & $\pm 0.7$
 & $-$ \\
${\cal B}(\eta \to \pi^+\pi^-\pi^0)$ &$-$ & $\pm 1.8$ &$-$ & $\pm 1.8$ &
$-$ & $\pm 1.8$ \\ 
$N_{B}$ & $\pm 1.3$ & $\pm 1.3$ & $\pm 1.3$ & $\pm 1.3$ & $\pm 1.3$ & $\pm 1.3$
\\ \hline
Sum & $^{+5.4}_{-5.6}$ & $^{+6.5}_{-6.6}$ & $^{+7.7}_{-6.4}$ & $^{+8.5}_{-8.7}$
    & $^{+8.9}_{-9.1}$ &$\pm 9.8$ \\ \hline     
 
\end{tabular}
\label{tab:syst}
\end{table*}

The possible detector bias due to the tracking acceptance  
for $A_{CP}(B^\pm\to \eta \pi^\pm)$ and $A_{CP}(B^\pm\to
\eta K^\pm)$ is evaluated using the $A_{CP}$ value of the continuum component.
No obvious bias is observed  and we use the  statistical error 
of the $\etagg$ and $\etapi$ modes combined as the systematic error.
 The bias error of 0.01
is added in quadrature with the fit systematic error to give the final
systematic uncertainty in Table I. 
Figures ~\ref{fig:acppi} and~\ref{fig:acpk} show the $\Mbc$ and $\de$ projections for the $B^+$ and $B^-$ samples. In both
the $\eta \pi^\pm$ and $\eta K^\pm$ modes, we observe larger $B^+$ signals.

\begin{figure}[htb]
\includegraphics[width=0.44\textwidth]{fig2.epsi}

\caption{$\Mbc$ and $\de$ projections for (left) $B^-\to\eta \pi^-$ and  
(right) $B^+ \to \eta\pi^+$ with 
the $\etagg$ and $\etapi$ modes combined. Open
histograms are data, solid curves are the fit functions, dashed lines show
the continuum contributions and shaded histograms are the contributions
from charmless $B$ decays.  The small contributions near $\Mbc = 5.28$ GeV/$c^2$ and 
$\de = -0.05$ GeV are the backgrounds from misidentified $\btoetak$ 
(reflections).}  
\label{fig:acppi}
\end{figure}

\begin{figure}[htb]
\includegraphics[width=0.44\textwidth]{fig3.epsi}

\caption{$\Mbc$ and $\de$ projections for (left) $B^-\to\eta K^-$ and  
(right) $B^+ \to \eta K^+$ with 
the $\etagg$ and $\etapi$ modes combined. All symbols are the same as in
 Fig. 2.  
The small contributions near 
$\Mbc = 5.28$ GeV/$c^2$ and 
$\de = 0.05$ GeV are the backgrounds from misidentified $\btoetapi$ 
(reflections). }  
\label{fig:acpk}
\end{figure}

The systematic error of the efficiency arises from the $\cal{R}$ requirement, 
tracking efficiency, particle identification, $K^0_S$ reconstruction, 
$\pi^0$ and $\eta_{\gamma\gamma}$ reconstruction, and $\etagg$ and $\etapi$
branching fractions.
The performance of the $\LR$ requirement is studied by 
checking the data-MC efficiency ratio using the $B^+\to \overline{D}{}^0 \pi^+$ 
control sample. The
 systematic errors on the charged track reconstruction
are estimated to be 1\% per track using  partially
reconstructed $D^*$ events.
 The $\pi^0$ and $\eta_{\gamma\gamma}$ 
reconstruction efficiency is verified by comparing the $\pi^0$
decay angular distribution with the MC prediction, and by measuring the ratio 
of the branching fractions for the two $D$ decay channels 
$\overline{D}{}^0 \to K^+\pi^-$
and $\overline{D}{}^0 \to K^+\pi^-\pi^0$.  
The $K_S^0$ reconstruction is verified by 
comparing the  ratio of $D^+\to K_S^0\pi^+$ and $D^+\to K^-\pi^+\pi^+$ yields. 
The uncertainties in the $\etagg$ and $\etapi$ branching fractions are taken 
from Ref.~\cite{pdg}.
Table ~\ref{tab:syst} summarizes the systematic uncertainties, including the 
error on the number of $\bb$ events. The systematic error that arises from 
how well PHOTOS describes  final state radiation 
is found to be negligible \cite{photoserr}. The final systematic error 
for  the combined branching fraction is obtained by assuming that the 
systematic errors for the sub-decay modes are 100\% correlated.  

 We  observe an excess of $B^0\to \eta K^0$ events but
the significance is slightly less than 3.  Therefore, an upper limit
on the branching fraction  at  90\% confidence level is provided. 
To calculate this limit, we find the yield for which 90\% of the area of the
likelihood function lies at lower values. We divide the yield by 
the reconstruction efficiency reduced by  $1 \sigma$ of its  
uncertainty, which is the quadratic sum of the errors given in Rows 2-8
of Table 2. The result is then inflated by the $1 \sigma$   
 uncertainty due to the parameters fixed in the fit (1st
row of Table 2) to obtain the upper limit including all systematic 
uncertainties.  
   
In summary, we have  observed  $\btoetapi$ and  
$\btoetak$ decays; their branching fractions are measured to be 
$(4.2\pm 0.4 \pm 0.2)\times 10^{-6}$ and $(1.9\pm  0.3^{+0.2}_{-0.1}) \times 
10^{-6}$, respectively. These results are consistent with our previously 
published measurements \cite{etahbelle} with statistical errors reduced by more than 40\%. Compared with the earlier BaBar results \cite{etahbabar}, 
our measurements are more precise despite  a $1.8 \sigma$ lower branching 
fraction on $B^\pm \to \eta K^\pm$. The $CP$-violating asymmetries are measured
to be $A_{CP}(B^\pm\to \eta \pi^\pm)=-0.23\pm 0.09 \pm 0.02$ and 
$A_{CP}(B^\pm\to \eta K^\pm)= -0.39\pm 0.16\pm 0.03$, which are $2.5 \sigma$ 
and $2.4 \sigma$ away from zero, respectively. It is interesting to note that
the $A_{CP}$ values for these two modes obtained by the BaBar collaboration 
are also negative, slightly more than $1 \sigma$ away from zero for each mode. 
Larger data  samples are  needed to verify these large $CP$ asymmetries. 
Finally, we find a hint of an $\eta K^0$ signal with  
 $\mathcal{B}(B^0\to \eta K^0) = 
(1.1 \pm 0.4 \pm 0.1)\times 10^{-6}$. Since the measurement is not significant, 
we provide an upper limit at the 90\% confidence level of 
$1.9 \times 10^{-6}$. A similar hint was also observed by the BaBar 
collaboration with a  central value of $(1.5 \pm 0.7 \pm 0.1)\times 10^{-6}$.
The combined average, $(1.2\pm 0.4)\times 10^{-6}$,  shows $3.4 \sigma$
evidence for the $CP$ eigenstate decay $B^0\to \eta K^0$.

We thank the KEKB group for the excellent operation of the
accelerator, the KEK Cryogenics group for the efficient
operation of the solenoid, and the KEK computer group and
the NII for valuable computing and Super-SINET network
support.  We acknowledge support from MEXT and JSPS (Japan);
ARC and DEST (Australia); NSFC (contract No.~10175071,
China); DST (India); the BK21 program of MOEHRD and the CHEP
SRC program of KOSEF (Korea); KBN (contract No.~2P03B 01324,
Poland); MIST (Russia); MESS (Slovenia); NSC and MOE
(Taiwan); and DOE (USA).

\end{document}